# Tuning light matter interaction in magnetic nanofluid based field induced photonic crystal-glass structure by controlling optical path length

Junaid M. Laskar[a]*, Baldev Raj[b] and John Philip[c]

[a]Max Planck Institute for Dynamics and Self-Organization, Göttingen 37077, Germany

[b]National Institute of Advanced Studies, IISC Campus, Bangalore 560 012, India

[c]SMARTS, MMG, Indira Gandhi Centre for Atomic Research, Kalpakkam 603102, T.N, India

*Corresponding author Email: junaid.laskar@ds.mpg.de

**Abstract**

The ability to control the light matter interaction and simultaneous tuning of both structural order and disorder in materials, although are important in photonics, but still remain as major challenges. In this paper, we show that optical path length dictates light-matter interaction in the same crystal structure formed by the ordering of magnetic nanoparticle self-assembled columns inside magnetic nanofluid under applied field. When the optical path length (L=80 μm) is shorter than the optical (for wavelength, λ=632.8 nm) coherence length inside the magnetic nanofluid under applied field, a Debye diffraction ring pattern is observed; while for longer path length (L= 1mm), a corona ring of scattered light is observed. Analysis of Debye diffraction ring pattern suggests the formation of 3D hexagonal crystal structure, where the longitudinal and lateral inter-column spacings are 5.281 and 7.344 microns, respectively. Observation of speckles within the Debye diffraction pattern confirms the presence of certain degree of structural disorder within the crystal structure, which can be tuned by controlling the applied field strength, nanoparticle size and particle volume fraction. Our results provide a new approach to develop next generation of tunable photonic devices, based on simultaneous harnessing of the properties of disordered photonic glass and 3D photonic crystal.

# 1. Introduction

Colloidal particles can self-assemble into various ordered periodic structures under favorable conditions (e.g. screened columbic repulsive interactions, volume fractions, pH, temperature etc.), which makes colloid a model soft condensed matter system to study the crystallization thermodynamics of atomic and molecular systems[1-6]. In order to study colloidal crystals, visible light diffraction has been routinely used as a diagnostic tool, since the length scale of the colloidal crystal is of the order of visible wavelength[7-10]. It also enables a potential route for the fabrication of three dimensional (3D) photonic crystals that can be used to control and manipulate light in three dimensions[11-15]. The disorder due to structural artifacts present in any photonic crystal gives rise to multiple scattering, which limits their technological applications[16]. To overcome this issue, the recent focus is on harnessing the multiple scattering, in order to introduce new functionalities, which opens up a completely new perspective on disorder in photonic crystals, e.g. random lasers employing photonic glasses containing a completely disordered arrangement of monodisperse colloids[16-19]. Hence, reversible tuning of disordered colloidal particles (photonic glass) into a 3D photonic crystal or vice versa with an external stimulus will have huge technological implication.

Magnetic nanofluid, popularly known as ferrofluid, is a promising colloidal system that offers interesting optical and photonic properties, due to their ability to reversibly change the internal structural ordering under an applied magnetic field[20-29]. The magnetic nanofluid is a stable dispersion of magnetic nanoparticles in a carrier liquid, where nanoparticles undergo random Brownian motion in the absence of any external field. On applying external magnetic field $(H_0)$, the magnetic nanoparticles in the dispersion acquire dipole moment $(m)$ given by

$$m = \frac{\pi}{6} d^3 \chi H_0 \tag{1}$$

Here, $d$ is the particle diameter, $\chi$ is the effective susceptibility of an individual nanoparticle. The diploar interaction energy $U_{ij}$ between two magnetic nanoparticles is given by

$$U_{ij}(r_{ij}, \theta_{ij}) = \frac{m^2 \mu_0}{4\pi} \left( \frac{1 - 3\cos^2 \theta_{ij}}{r_{ij}^3} \right) \tag{2}$$

where $\mu_0$ is the magnetic permeability of free space, $r_{ij}$ is the magnitude of the vector describing the distance between the centers of $i^{th}$ and $j^{th}$ nanoparticles, and $\theta_{ij}$ is the angle between the vector $r_{ij}$ and the external magnetic field vector. The effective interaction energy between two magnetic nanoparticles is determined by the competition between the dipolar interaction energy $U_{ij}(r_{ij},\theta_{ij})$ and the thermal energy $(k_B T)$, hence, can be described by

$$\Lambda = \frac{U_{ij}(r_{ij},\theta_{ij})}{k_B T} \tag{3}$$

Here, $k_B$ is the Boltzmann constant and $T$ is the temperature. When $U_{ij}(r_{ij},\theta_{ij})$ becomes at least one order of magnitude greater than $k_B T$, i.e. when $\Lambda > 1$, linear aggregates, i.e. chains are formed due to the self-assembly of nanoparticles along the applied field direction. On further increase in field strength $(H_0)$, the chains of nanoparticles undergo lateral aggregation, known as zippering aggregation, due to the nanoparticle chain fluctuation induced short range attraction to minimize the total dipolar potential energy of the system[30]. Therefore, the application of external field leads to complex structural changes inside the magnetic nanofluid that gives rise to interesting optical properties. As the magnetic field increases, the scatterer size increases from single nanoparticle diameter to the dimensions of columns of nanoparticles, which lead to an increase in the scattering efficiency, thereby resulting in a decrease in the transmitted light intensity[31]. However, for a given volume fraction $(\phi)$ and applied field strength $(H_0)$, the transmitted light intensity decreases on increasing the optical path length $(L)$[32]. When the scatterer shape changes from spherical (single nanoparticles) to the cylindrical shape (columns), the scattered pattern changes from a single laser spot to a corona like ring, with the transmitted laser spot on the ring circumference, for a longer path length $(L = 1mm)$ [31,33]. However, for very short path length $(L \leq 10 \mu m)$, the interaction of the white light with the structural periodicity due to two dimensional (2D) hexagonal crystal ordering results in chromatic diffraction rings[23,34,35].

The path length dependent field induced column spacing is studied in oil-in-water ferrofluid emulsions of relatively larger oil droplet size (>100 nm)[36]. The field induced crystal ordering is observed to be 2D hexagonal in magnetic nanofluid with small particle size (~ 11.8 nm) for very short path lengths $L(\leq 10\mu m)$ [34]. On increasing the path length to $L(=0.5mm)$ for magnetic nanofluid with larger core-shell structure nanoparticles (>100nm), the field induced crystal ordering becomes 3D hexagonal[22]. However, the field induced crystal structure for the intermediate path length range, i.e. $10\mu m < L < 0.5mm$ and the spacing among the columns of nanoparticles for longer path lengths $(L > 10\mu m)$ in magnetic nanofluids are still unknown ; owing to experimental constrains and lack of stable suspensions with long term stability, which is a prerequisite to observe reversibly tunable optical properties. As a result, an understanding of the role of path length on the magnetic field induced crystal ordering and the resulting optical properties such as diffraction, light transmission, scattering, speckle pattern, Fano resonance etc. are still missing [23,31,37-45].

Moreover, the structural studies carried out so far have been limited to magnetic nanofluids of relatively larger magnetic particle size (diameter $d > 100nm$) and higher concentration (volume fraction, $\phi > 0.1$), i.e. shorter inter-particle distance ($r_{ij}$), which leads to a stronger dipolar interaction $(U_{ij})$ [Eq.(2)] among the magnetic nanoparticles[22]. This results in the sedimentation of particles even in the absence of external field, and thereby limiting the possibility to tune colloidal crystalline order to create photonic crystals[46]. In this study, we aim to probe the optical path length dependent diffraction and scattering pattern in a stable colloidal suspension of $Fe_3O_4$ nanoparticles to explore the possibility of achieving a reversibly tunable colloidal photonic glass to a 3D photonic crystal transition.

**2. Material and method**

**2.1. Magnetic nanofluid sample**

We synthesized a highly monodisperse (polydispersity index, PDI~ 0.092) and stable magnetic nanofluid which consists of very small superpramagnetic magnetite ($Fe_3O_4$) particles (diameter, $d \sim 6.7nm$) with a surfactant layer (oleic acid, layer thickness ~ 1.5 nm) for the present study.[31,38] The volume fraction used for the nanoparticles dispersed in the carrier liquid kerosene was very low $(\phi = 0.01713)$.

## 2.2. Experimental section

A stabilized He-Ne laser (Spectra-physics,117A) of wavelength $\lambda \sim 632.8 nm$ with an output power ~ 4.5 mW (coherence length > 100 m) is used as a light source. An external magnetic field $(H_0 \sim 300G)$ is applied along the incident light direction by keeping the cell containing the magnetic nanofluid sample inside a solenoid[31]. The light scattered patterns are acquired on a translucent screen, kept at a distance of 11.8 cm from the sample cell by using a digital single lens reflex (DSLR) camera, for two optical path lengths $(L = 1mm \& 80 \mu m)$ of the sample cell. Details of the experimental technique and schematic of the experimental set up are described elsewhere[31,38].

## 3. Results and discussion
### 3.1. Observed light scattered and diffraction pattern

For the longer optical path length $(L = 1mm)$, a single corona ring with the transmitted laser spot on the ring circumference is observed in the scattered pattern, as shown in Figure 1(a). However, for the shorter path length $(L = 80 \mu m)$, concentric Debye-Scherrer diffraction rings with the transmitted laser spot at the center is observed in the scattered pattern, as shown in Figure 1(b). Although, the complete ring circumference is visible for the innermost diffraction ring, the light intensity and the visible radial angle range of the ring circumference gradually decreases for the outer rings. Speckle patterns are observed along the broad circumferential widths of the diffraction rings and in the intermediate space among the rings. The innermost and the other three concentric outer rings are designated as 1$^{st}$, 2$^{nd}$, 3$^{rd}$ and 4$^{th}$ ring for the analysis purpose.

### 3.2. Origin of single corona like scattered ring and the concentric diffraction ring pattern

The observed single corona like ring for the longer optical path length [Figure 1(a)] originate from the light scattering from the cylindrical surface of the columns of aggregated nanoparticles, formed along the applied field direction[31,33]. When the incident light interacts with a cylindrical surface, the resulting outgoing scattered light takes the shape of a scattered light cone. The intersection of this scattered light cone on a screen placed perpendicular to the incident light gives the observed scattered pattern, which forms a conic section. When light is incident at near zero degree ($\zeta=0°$) with respect to the cylinder axis, the observed scattered pattern takes the shape of a circular ring with the transmitted light spot on the ring circumference. Since, the incident light is along the direction of the applied magnetic field, the observation of corona like ring on the screen confirms the formation of columns[31,33]. Since, the single corona ring arises due to the light scattering from the surface of cylindrical columns, it does not contain any information about the periodicity of the ordered structure formed by the columns. Moreover, the shape of the scattered light pattern, described by the conic section, depends on the incident angle ($\zeta$) of light with respect to the cylindrical axes of the columns[33]. The observed shape of the scattered pattern changes from a circular ring to a straight line as the incident angle changes from $\zeta=0°$ to $\zeta=90°$ as shown in Figure 2(a-c). This is yet another convincing evidence for the scattering of light from nanoscale cylindrical structures in the ferrofluid under a magnetic field.

Figure 2 (d-f) shows the microstructural evolution of the field induced structures at three different magnetic field strengths in a ferrofluid emulsion with a droplet size ~ 200 nm. Ferrofluid emulsion of this size is chosen in order to observe the microstructure using a simple phase contrast optical microscope. In the case of magnetic nanofluid, visualization of such structure formation is not possible with an optical microscope, owing to their small size. The increase in the length of the chains formed by the aggregation of nanoparticles is evident from the microscopic images.

Figure 2 (g & h) shows the transmitted light intensity of magnetic nanofluid $(\phi=0.01713)$, as a function of the applied magnetic field for different ramp rates, for $L=1mm$. For all the ramp rates of applied magnetic field, the transmitted intensity decreases due to the increase in Mie scattering efficiency, resulting from the increase in scatterer size, due to an increase in the dimensions of columns, on increasing the applied field strength[31]. By controlling the ramp rate of applied magnetic field, the field exposure time is tuned to change the nanoparticle aggregation rate. For a nanofluid under a given applied field strength, the required aggregation time of nanoparticles depends on the competition between the dipolar interaction energy and the viscous force experienced by the nanoparticle aggregates[32]. The required aggregation time decreases with the increasing field strength $(H_0)$. If the exposure time of the applied field is equal to the aggregation time, then the nanoparticle aggregation takes place at the fastest rate. The aggregation rate again slows down when the field exposure time is either more or less than the required aggregation time[32]. Therefore, for an optimum ramp rate (2.5 G/s), when the field exposure time is equal to the aggregation time, the rate of nanoparticle aggregation, i.e. the rate of increase in scatterer size is fastest, whereas, the aggregation rate again decreases gradually, either on increasing or decreasing the ramp rate from the optimum value, as shown in Figure 2(g & h). Therefore, the variation of transmitted light intensity through a magnetic nanofluid can be easily tuned by controlling the ramp rate of the applied magnetic field.

The observation of concentric Debye-Scherrer diffraction ring pattern [Figure 1(b)], for the cell with shorter path length $(L = 80 \mu m)$, originates from the interaction of light with field induced crystal structure, formed by the periodic spacing of the columns occupying lattice vertices, inside the nanofluid[23]. In order to observe the diffraction pattern from an ordered crystal structure, the scattered light emanating from a particular set of parallel lattice planes with a constant phase difference interfere constructively[47]. The constant phase difference among the scattered waves propagating through an optical media is maintained until the temporal coherence[48]. The coherence length in an optical media is a product of coherence time and the group velocity of the light wave in that media[49]. Therefore, the light diffraction phenomenon in an ordered structure can be observed, only if the wave propagation path length is shorter than the coherence length of an optical media. The coherence length of the stabilized He-Ne laser, used in this experiment is greater than 100 m in air and hence much more than the sample cell path lengths used $(L = 1mm \& 80 \mu m)$. However, it is well established that the coherence length drastically decreases when propagating through a highly scattering liquid media.[50] As the scatter size increases with increasing field strength in a magnetic nanofluid, it resembles a highly scattering liquid media. Therefore, the observations of Figure 1 confirm that in a magnetic nanofluid kept under an applied magnetic field, the value of the coherence length lies within the range of 80 μm and 1 mm.

**3.3. Analysis of magnetic field induced crystal ordering from the diffraction rings**

The observation of the diffraction rings [Figure 1(b)] confirms the formation of a crystalline structure by the field induced columns. The continuous circumferences of the 1st and 2nd rings indicate that the formed structure is polycrystalline, irrespective of the type of crystal lattice. The broader width of the ring, unlike the sharp diffraction rings for atomic crystals, and the speckle pattern within the diffraction pattern indicate that the crystal ordering is not perfect, but has a certain degree of structural disorder[51,52].

In order to find out the exact crystal structure in the present case $(L = 80 \mu m)$, as a first approximation, the field induced crystal structure is assumed to be a 2D hexagonal, as observed earlier for a very short path length $(L \leq 10 \mu m)$ [23,34]. A 2D crystal acts as a two dimensional diffraction grating and the resulting diffraction rings pattern follows [10,51]

$$d_{hk} \sin \theta = n\lambda \qquad (4)$$

where $\lambda$ is the incident light wavelength, $n$ is the diffraction order (an integer), $d_{hk}$ is the spacing between the parallel lattice planes, more appropriately parallel lattice lines for 2D crystals designated by the Miller indices $(hk)$. Here, the reflection angle $(\theta)$ is the angle between the incident light and the lattice planes. Therefore, each diffraction ring, observed at a particular diffraction angle $(\theta_d)$, results from the constructive interference of the light scattered from a particular family of parallel lattice planes $(hk)$ of the crystal structure.

The diffraction angle $(\theta_d)$ corresponding to each Debye-Scherrer ring of the observed diffraction pattern [Figure 1 (b)] is determined by measuring the sample to screen distance (11.8 cm) and the ring radius. For this purpose, the light intensity of each Debye-Scherrer ring is integrated over the entire radial angle [0°-360°] by considering the transmitted laser spot as the origin, and plotted as a function of $\theta_d$, as shown in Figure 3(a). However, it is observed that the broad diffraction intensity peaks are asymmetric in shape and the positions of the maximum light intensity are shifted towards a lower diffraction angle $(\theta_d)$ for each peak. This is due to the presence of a background formed by scattered light that is observed within the diffraction pattern [Figure 1(b), also shown in the inset of Figure 3(b)]. Since, the field induced column dimensions are of the order of the incident light wavelength ($\lambda \sim$ 632.8 nm); significant scattering of light occur, which forms the background within the observed diffraction pattern [Figure 1(b)][31]. Moreover, the decrease in the intensity of the scattered light with increasing scattering angle results in a decrease in the background light intensity with increasing $\theta_d$, thereby leading to the shift of maximum intensity positions towards lower diffraction angle values $(\theta_d)$ in each peak, as observed in Figure 3(b)[53]. This background scattered light contributes to the finite light intensity in the dark regions among the bright diffraction rings [Figure 3(a)].

Therefore, the scattered light background is subtracted from the observed diffraction pattern [Figure 1(b)] in order to correct the maximum intensity positions [diffraction angles $(\theta_d)$] of the diffraction intensity peaks. The diffraction peak shapes remain symmetric after the background corrections, as shown in Figure 3(b). The inset of Figure 3(b) shows the diffraction pattern after subtracting the scattered light background. The center positions for the first three diffraction intensity peaks are determined by Gaussian fit. The obtained peak center positions also coincided very well with the center of the circumferential widths of the concentric Debye-Scherrer rings. The diffraction angle $(\theta_d)$ corresponding to the fourth ring is determined by positioning a circle along the center of ring circumferential width [Inset of Figure 3(b)]. Because of the very low light intensity and small radial angle range [~ 0°-45°] over which the intensity is integrated, the peak shape still remains highly asymmetric [Figure 3(b)]. Therefore, the Gaussian fit is not possible. The obtained diffraction angle $(\theta_d)$ values corresponding to the four Debye-Scherrer diffraction rings are shown in the Table. 1.

The inter-planar spacing value $(d_{hk})$, as shown in Table 1, is determined by using Eq. (4) and the reflection angle $(\theta)$, which is equal to the diffraction angle $(\theta_d)$ of the corresponding ring for a 2D crystal. For a 2D hexagonal lattice, also known as the triangular lattice, the inter-planar spacing $(d_{hk})$ as a function of Miller indices $(hk)$ and primitive lattice vector $a$ is given by [51]

$$\frac{1}{d_{hk}^2} = \frac{4}{3}\left(\frac{h^2 + hk + k^2}{a^2}\right) \qquad (5)$$

If the assumed crystal lattice structure (2D hexagonal) and the Miller indices $(hk)$ are correct, then all the $d_{hk}$ values corresponding to each diffraction ring should produce the same value of $a$, on using Eq. (5). In this regard, nearly the same value of $a$ is obtained on considering the two most basic 2D hexagonal Miller planes, $(10)$ and $(11)$ for the 1$^{st}$ and 2$^{nd}$ rings, respectively.

Therefore, 2D hexagonal ordering is still present along the two directions perpendicular to the applied field direction, for $L = 80 \mu m$, as observed earlier for very short path length $(L \leq 10 \mu m)$ [34]. Similar analysis carried out for the 3$^{rd}$ and 4$^{th}$ rings, on considering the next most probable 2D hexagonal crystal Miller planes, $(20)$ and $(21)$, respectively, gives different values of $a$, as shown in Table 1. In fact, no other Miller planes could be found to produce the same value of $a$ for the $d_{hk}$ values corresponding to the 3$^{rd}$ and 4$^{th}$ rings. Moreover, the measured diffraction angle values $(\theta_d)$ do not match with any other 2D crystal lattice structures, including the square and rectangular. This clearly indicates that the magnetic field induced crystal lattice is not 2D. As the 3$^{rd}$ and 4$^{th}$ rings must be due to the structural periodicity along the 3$^{rd}$ dimension, i.e. the applied field direction, the possibility of a three dimensional (3D) crystal structure is present. Such a 3D hexagonal field induced crystal structure is observed in a magnetic nanofluid with long path length $(L = 0.5 mm)$ and a larger size (> 100 nm) core-shell nanoparticles[21]. Therefore, for $L = 80 \mu m$, we consider the possibility of the formation of a 3D hexagonal crystal where the field induced columns occupy the vertices of the crystal lattice, as shown in Figure 4.

A 3D hexagonal crystal acts as a 3D diffraction grating and obeys Bragg's law[51]

$$2d_{hkl} \sin \theta = n\lambda \qquad (6)$$

where $d_{hkl}$ is the spacing between the parallel lattice planes i.e., Miller indices $(hkl)$ and $\theta$ is the angle between the incident light and the lattice planes. However, for a 3D crystal, the relation between diffraction angle $\theta_d$ and the reflection angle $\theta$ is $\theta_d = 2\theta$. The inter-planar spacing $d_{hkl}$ for a 3D hexagonal lattice is given by[51]

$$\frac{1}{d_{hkl}^2} = \frac{4}{3}\left(\frac{h^2 + hk + k^2}{a^2}\right) + \frac{l^2}{c^2} \qquad (7)$$

Here, $a$ and $c$ are the primitive lattice vectors. Physically, the primitive lattice vector $a$ signifies the lateral spacing among the columns, i.e. structural periodicity along the two directions perpendicular to the applied field; while $c$ signifies longitudinal spacing among the arrays of columns, which in other words is the longitudinal spacing among the columns formed along the applied field direction, as shown in Figure 4. The $1^{st}$ and $2^{nd}$ rings are assumed to be due to diffraction from the two most basic 3D hexagonal Miller planes, $(100)$ and $(110)$, and their corresponding equivalent planes, which are related by crystal symmetry. The value of the primitive lattice vector $a$ is found to be nearly the same for the $1^{st}$ and $2^{nd}$ rings, on using the Eqs. [(6) & (7)], as given in Table 2. The $3^{rd}$ and $4^{th}$ rings are assumed to be due to diffraction from the next most probable higher order Miller planes, $(201)$ and $(211)$, respectively. The value of lattice vector $c$ is obtained by analyzing the $3^{rd}$ ring (table 2), by using the average of the values of $a$ as obtained from the $1^{st}$ and $2^{nd}$ rings and the Eqs. [(6) & (7)]. Now, by using this value of $c$ ($3^{rd}$ ring), the value of lattice vector $a$ is obtained for the $4^{th}$ ring. Interestingly, this value of $a$ ($4^{th}$ ring) is found to be nearly the same as the average of the values of $a$, as obtained from $1^{st}$ and $2^{nd}$ rings. This observation of the nearly same values of the primitive lattice vectors $(a\ \&\ c)$, for all the four diffraction rings, further strengthen our argument of the formation of a 3D hexagonal lattice. The value of $c (= 5.28\ \mu m)$ and the average of the values of $a (= 7.34\ \mu m)$ for all the four diffraction rings are the average linear and lateral spacing values among the field induced columns, as shown in Figure 4.

The field induced crystal structure is a function of the sample path length ($L$, confinement thickness) along the applied magnetic field direction[21,34]. On increasing the magnetic field strength $(H_o)$, the magnetic dipolar interaction energy $(U_{ij})$, therefore, the coupling constant $(\Lambda)$ among the magnetic nanoparticles increases [Eqs. (2 & 3)]; which leads to the formation of linear chains along the applied field direction. On further increase in applied field strength, the chains of nanoparticles undergo aggregation, lateral to the direction of applied field, known as zippering aggregation, thereby forming columns; due to chain fluctuation induced short range attraction and the system requirement to minimize the total dipolar potential energy of the system[30]. For ferrofluid emulsion, magnetic field induced zippering aggregation among the chains is shown in Figure 2(f). For magnetic nanofluid, the summation of dipolar interaction energies, among all the particles constituting different columns, gives the total potential energy of the system under the applied field. However, the main contribution to the total potential energy comes from the dipolar interactions among the nanoparticles of the nearest neighbor columns, as the dipolar interaction strength decreases rapidly with increasing inter-particle distance $(r_{ij})$ [Eq. (2)]. For a given sample volume, the minimization of total system potential energy is achieved by: (a) maximizing the inter-column spacing, therefore, the inter-particle distance $(r_{ij})$ and (b) by choosing the structural arrangement of columns, which maximizes the possibility of $\theta_{ij}$ to approach $0°$ or $180°$, which results in negative values of $U_{ij}$, i.e. lower potential energy [Eq. (2)]. For shorter sample path length, e.g. $L \leq 10\mu m$, the 2D hexagonal periodic spacing among the columns lateral to the applied field direction maximizes the inter-column spacing and hence minimizes the system potential energy [Eq. (2)][34]. Moreover, on increasing the widths of columns by zippering, the columns become rigid, which lead to the dominance of long range end pole repulsion among the columns over the short-range fluctuation induced attraction of the chains of nanoparticles[35]. Therefore, in order to maximize the inter-column spacing and due to the end pole repulsions among the rigid columns, 2D hexagonal crystal structure is formed for the short path lengths of magnetic nanofluid under applied field[34].

However, simultaneous increase in column length with the increase in path length will also increase the number of particles with $\theta_{ij} = 90°$ [$U_{ij} = +ve$, as per Eq. (2)], known as the in-registry positions for two neighboring columns, which will again result in the increase in the potential energy of the system[30]. Therefore, for the longer sample path length ($L = 80\mu m$), 2D hexagonal lattice ordering is avoided by the formation of 3D hexagonal structure; where the column lengths are shorter and has a lower system potential energy due to the reasons mentioned below. For a given sample volume with a longer path length ($L = 80\mu m$), the number of particles with $\theta_{ij}$ approaching $0°$ or $180°$ for two neighboring columns increases significantly for 3D hexagonal lattice, although the inter-column spacing, $r_{ij}$ decreases. The increase in the number of particles with $\theta_{ij}$ approaching to $0°$ or $180°$ and the simultaneous decrease in the number of particles with $\theta_{ij}(=90°)$ make $U_{ij}$ negative [Eq. (2)]. The dominant $U_{ij}$ contribution over the decrease in $r_{ij}$ leads to a decreasing the total potential energy of the system. This is the reason why the magnetic field induced crystal lattice spacing in magnetic nanofluid, formed by the spacing of columns, undergoes transition from 2D hexagonal to 3D hexagonal on increasing the path length from $L \leq 10 \mu m$ to $L = 80 \mu m$ [22,34].

The Miller planes corresponding to each diffraction ring appear dissimilar, despite being crystallographically equivalent due to structural symmetry, as shown in Table 2. In order to address this issue, the Miller-Bravais $(hkil)$ indexing is usually used for 3D hexagonal crystal, where a crystallographic plane is expressed in the form of four basic vectors $a_1, a_2, a_3$ and $c$, as shown in Figure5 . The index $i$ is the reciprocal of the fractional intercept on the $a_3$ axis. The value of $i$ depends on the values of $h$ and $k$ by the following relation $h+k=-i$ [51]. The advantage of using $(hkil)$ representation is that it can give similar indices to similar planes. The equivalent planes, which are related by symmetry are called *planes of a form*, and when enclosed in braces $\{hkil\}$, stands for the whole set, as given in Table 2. One representative Miller-Bravais plane $(hkil)$ corresponding to each diffraction ring is shown in the schematic of 3D hexagonal lattice in Figure 5 for the visualization purpose.

**3.4. Implication of magnetic field induced disorder present in the ordered 3D crystal structure.** The degree of crystal ordering inside the magnetic nanofluid is a function of the applied field strength $(H_0)$, which influences the dipolar interaction energy $U_{ij}(r_{ij}, \theta_{ij})$ and the coupling constant $(\Lambda)$ [Eqs. (2 & 3)]. As the maximum applied field strength $(H_0)$ in the present experimental geometry is ~ 300 G, the dipolar interaction energy $U_{ij}(r_{ij}, \theta_{ij})$ among the magnetic nanoparticles could not be increased significantly. Further, the very small particle size $(d)$ and lower volume fraction $(\phi)$, i.e. large inter-particle spacing $(r_{ij})$ also restricted to the maximum dipolar interaction energy $U_{ij}(r_{ij}, \theta_{ij})$. Therefore, both the positions and orientations of the field induced columns, forming the vertices of the crystal lattice, are distributed over a broad range as compared to a perfect 3D hexagonal crystal. Due to the above mentioned experimental conditions, imperfections and disorder are present in the crystal structure; which makes the structure polycrystalline, as evident from the observation of complete circumferential diffraction rings (1$^{st}$ and 2$^{nd}$), as shown in Figure 1(b). However, the degree of structural disorder and polycrystallinity can be tuned further by controlling the magnetic field strength $(H_0)$, particle size, monodispersity of dispersed particles and the nanofluid volume fraction $(\phi)$.

Now, for a given family of Miller planes $(hkl)$, a broad range of inter-planar spacing $(d_{hkl})$ values satisfy the diffraction condition [Eq. (6)], for a range of diffraction angle $(\theta_d)$ values, resulting in the observation of Debye-Scherrer diffraction rings with a broad circumferential width [Figure 1(b)]. The broad range of $d_{hkl}$ values also leads to small variation in the values of primitive lattice vectors $(a\ \&\ c)$ for different Debye-Scherrer diffraction rings (Table 2), obtained by using Eqs. (6 & 7). The disorder induced imperfection in crystal ordering does not allow the diffraction condition to be satisfied completely for higher order crystal lattice planes, which seems to be the reason for the observation of a small radial angle range of the visible circumference of the corresponding Debye-Scherrer diffraction rings (3$^{rd}$ and 4$^{th}$), as shown in Figure 1(b). Moreover, the disorder present in the ordered crystal structure results in multiple light scattering, which gives rise to the speckle pattern observed within the diffraction ring pattern.

Therefore, the structural configuration observed in the ferrofluid mimics a photonic glass integrated with a 3D photonic crystal[19,40,52,54,55]. Photonic glass, a medium containing a completely disordered arrangement of monodisperse colloids and characterized by the speckle pattern, showed random lasing originating from multiple scattering[19]. In photonic glasses, the Mie resonances originating from the constituting colloidal particle size (~ visible light wavelength) and the particle refractive index contrast with the medium strongly influence the light transport and the random lasing phenomenon[56]. In contrast, for magnetic nanofluid based photonic glass integrated with photonic crystal, Mie resonances originating from the magnetic field induced columns ( column size ~ visible light wavelength) and the high refractive index contrast of nanoparticles [$Fe_3O_4$ (2.42)] with the carrier liquid [kerosene (1.44)] induces weak localization of light and therefore, influences the light transport behavior[31,38]. Fano resonance, originating from the interference of multiple scattering light pathways between the Mie and Bragg scattered lights, is theoretically shown to govern the light transport for a photonic crystal with a certain degree of structural disorder[57,58]. Simultaneous presence of both structural order and disorder in the same system also gives rise to both light localization and Fano resonance in magnetic nanofluids[38,41]. Our finding of photonic glass property with 3D photonic crystalline behavior of magnetic nanofluid under applied magnetic field supports the recent theoretical finding that Fano resonance and light localization of the same physical origin, which results from the interference among different scattering paths arising from the multiple light scattering by the structural disorder [38,41,58].

## 4. Conclusions

We demonstrated that the optical path length $(L)$ dictates the light matter interaction (diffraction or scattering pattern) in a stable colloidal suspension of magnetic nanoparticles, known as the magnetic nanofluid, under applied magnetic field. We observed a single corona like light scattered ring for the longer optical path length $(L = 1 mm)$, due to the scattering of light from the surface of cylindrical columns of nanoparticle aggregates formed along the applied field direction; whereas we observed concentric diffraction (Debye-Scherrer) ring pattern for shorter optical path length $(L = 80 \mu m)$, since it is less than the optical coherence length of magnetic nanofluid under applied field. The analysis of the diffraction rings suggests the formation of 3D hexagonal crystal structure inside the magnetic nanofluid under applied field, which confirms that the transformation of the field induced crystal structure from 2D hexagonal to 3D hexagonal takes place in the range $10 \mu m < L < 80 \mu m$; therefore, explains the path length dependent field induced crystal ordering inside magnetic nanofluid[22,34]. The analysis of the Debye diffraction rings also suggests that the spacing between the columns in the lateral and longitudinal directions are 7.344 µm and 5.281 µm, respectively. To the best of our knowledge, this is the first experimental observation of a 3D colloidal crystal, which is formed by the non-contact spatial ordering of columns of aggregated nanoparticles, not by the packing of particles; thereby making the crystal periodicity much larger than the individual colloidal particle diameter [13,59]. Moreover, the observation of speckles within the Debye diffraction ring pattern confirms the presence of certain degree of structural disorder embedded within the 3D crystal structure, making it photonic glass integrated with photonic crystal; where the degree of ordering can be tuned by controlling the applied magnetic field strength, nanoparticle size and particle volume fraction. Our results showing the ability to reversibly tune and simultaneously harness the degree of both the structural disorder and order in magnetic nanofluid based photonic materials using an external stimulus open up the possibility to develop new generation of tunable photonic devices, e.g. tunable random lasers[16,19,60].

## Acknowledgements

The authors thank Prof. V. A. Raghunathan of Raman Research Institute Bangalore for stimulating discussions and Mr. P. Sukumar for experimental assistance. J.P thank Dr. T. Jayakumar and Dr. P. R. Vasudeva Rao for support. J.P. thanks the Board of Research in Nuclear Sciences (BRNS) for support through a research grant for the advanced nanofluid development program.

**Figure captions**

Figure 1  Light scattered patterns of magnetic nanofluid $(\phi=0.01713, d=6.7nm)$ under applied magnetic field $(H_0 \sim 300G)$ for two optical path lengths $(L)$ (a) Single corona ring for $L=1mm$ and (b) Debye-Scherrer diffraction ring pattern for $L=80\mu m$. The (c) and (d) are the 3D surface intensity distribution plots of (a) and (b), respectively. The direction of the applied magnetic field is along the incident light propagating through the optical path of the cell.

Figure 2  (a-c) Scattered light patterns of the ferrofluid $(\phi=0.00819)$ for different incident angles (ζ) of light with respect to the applied magnetic field direction for optical path length $L=1mm$, (d-f) microstructural evolution of the field induced structures for three different magnetic field strength in a ferrofluid emulsion with droplet size ~ 200 nm. Note that ferrofluid emulsion is chosen for this purpose in order to observe the microstructure with a simple phase contrast optical microscope. In the case of ferrofluid, visualization of such structure formation was not possible with an optical microscope owing to their small size. The increase in the aspect ratio of the column length is evident from the microscopic images. (f-g) Transmitted light intensity as a function of applied magnetic field for different ramp rates of ferrofluid $(\phi=0.01713)$ for $L=1mm$.

Figure 3  Intensity, integrated over the entire radial angle [0-2π] of the diffraction pattern (shown in the inset) by considering the transmitted laser spot as the origin, as a function of the diffraction angle $(\theta_d)$ for the (a) observed diffraction pattern and (b) after subtraction of the scattered background from the observed diffraction pattern.

Figure 4  Schematic of the 3D hexagonal ordering of field induced columns, formed by self-assembly of nanoparticles, inside a magnetic nanofluid, where the applied field and incident light directions are parallel to each other.

Figure 5  Visualization of one representative Miller-Bravais plane $(hkil)$ for each diffraction ring and indexing the diffraction rings with the whole set of equivalent lattice planes related by crystal symmetry using *planes of a form* $\{hkil\}$.

Table 1  Diffraction ring pattern analysis on considering 2D hexagonal crystal ordering.

Table 2  Diffraction ring pattern analysis on considering 3D hexagonal crystal ordering.

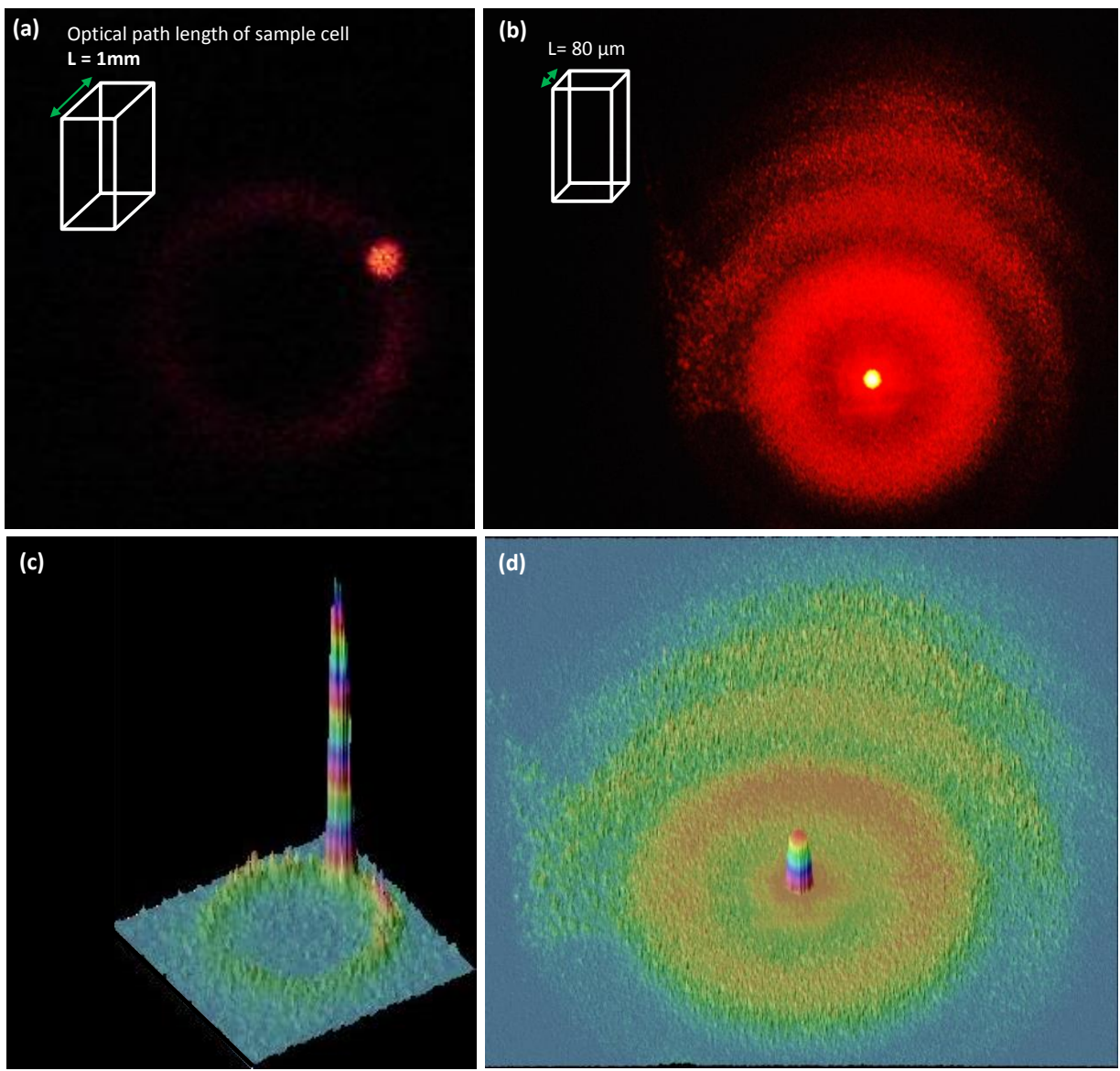

Figure 1

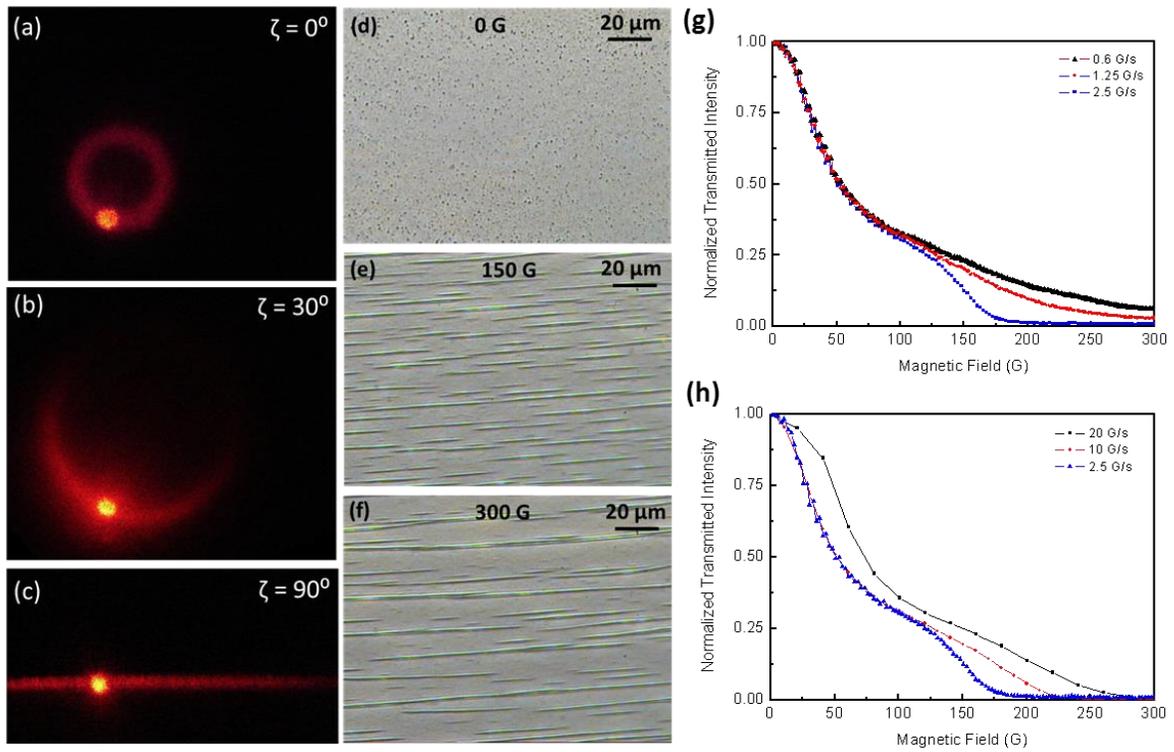

Figure 2

(a)

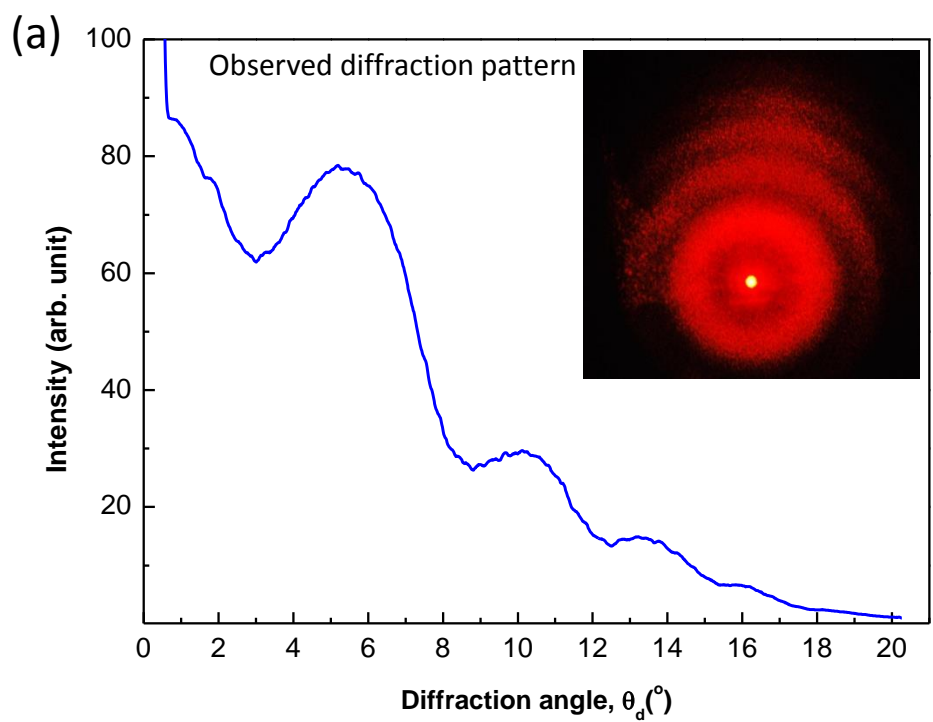

(b)

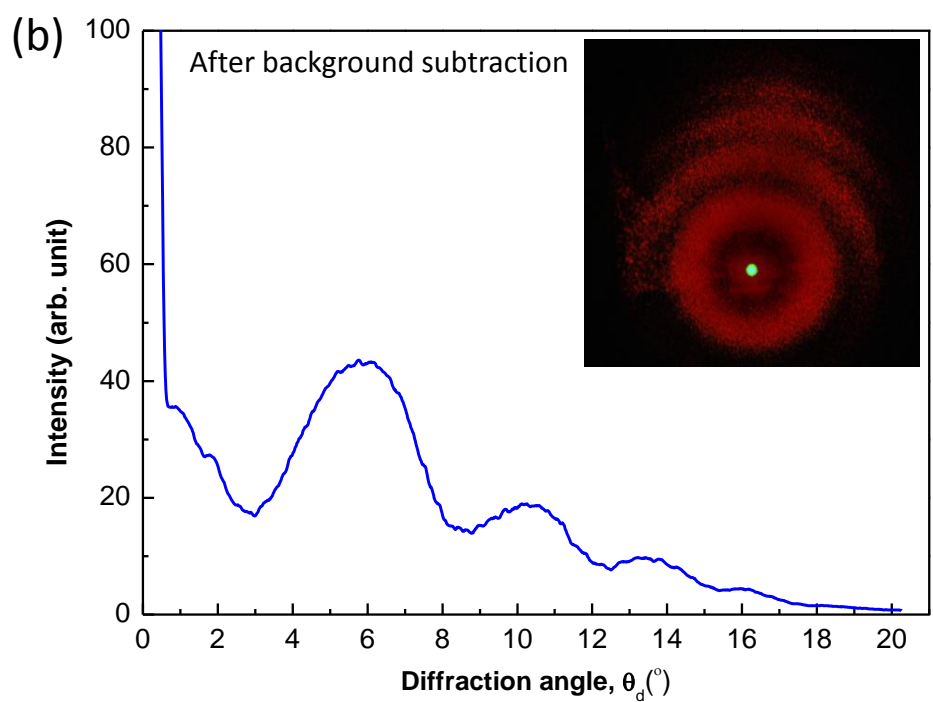

Figure 3

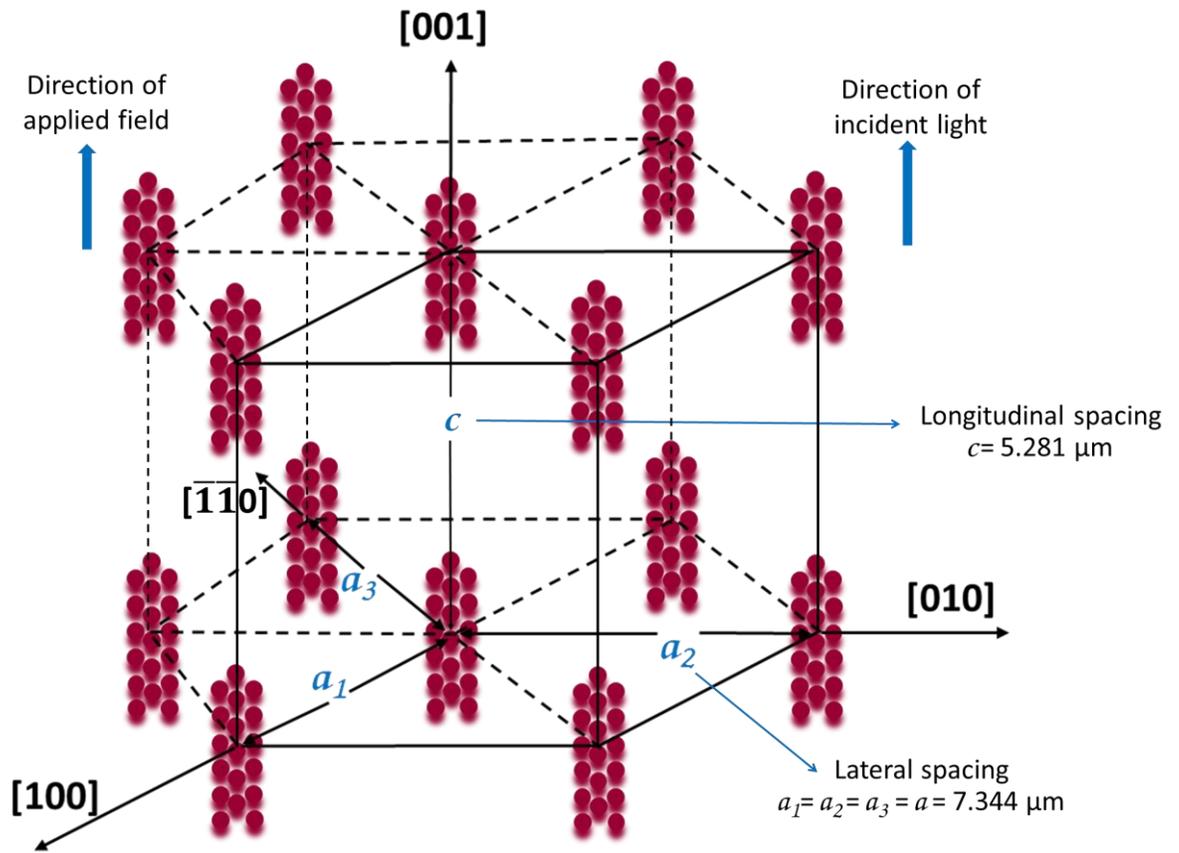

Figure 4

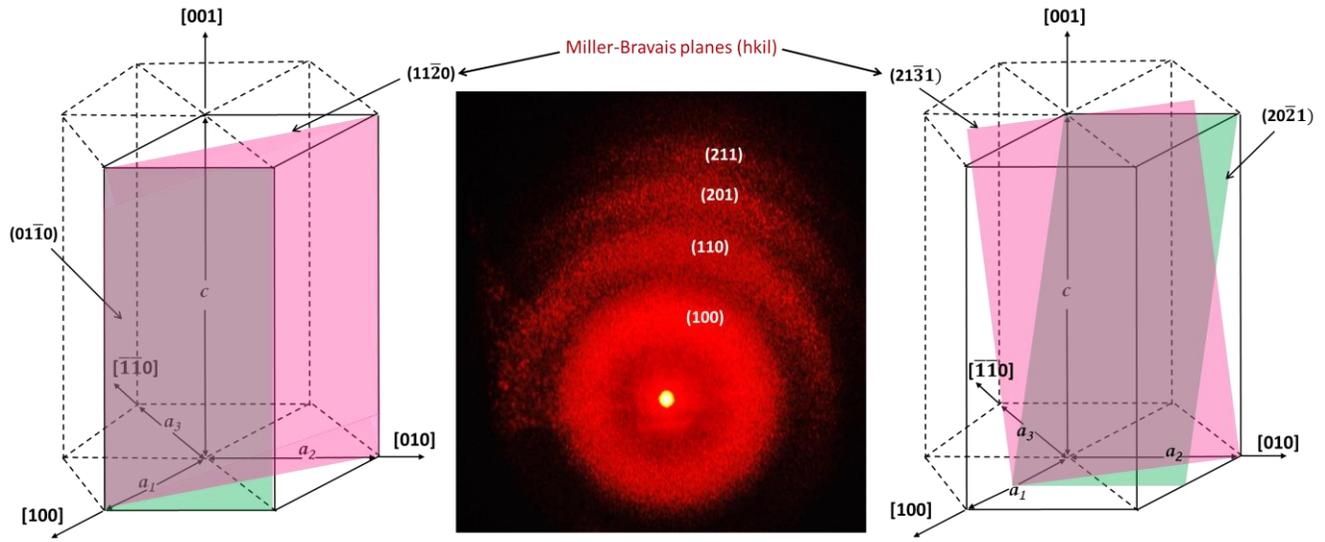

Figure 5

Table1 | Diffraction ring pattern analysis on considering 2D hexagonal crystal ordering

| Ring number | Diffraction angle $\theta_d$ (°) | Interplanar spacing $d_{hk}$ (μm) | Miller planes $(hk)$ | Primitive lattice vector $a$ (μm) |
|---|---|---|---|---|
| 1st | 5.701 | 6.370 | (10) | 7.356 |
| 2nd | 9.961 | 3.658 | (11) | 7.317 |
| 3rd | 13.374 | 2.736 | (20) | 6.318 |
| 4th | 16.486 | 2.230 | (21) | 6.812 |

Table 2 | Diffraction ring pattern analysis on considering 3D hexagonal crystal ordering

| Ring number | Miller planes $(hkl)$ | Miller-Bravais planes $(hkil)$ | Planes of a form $\{hkil\}$ | Primitive lattice vector | |
|---|---|---|---|---|---|
| | | | | $a$ (μm) | $c$ (μm) |
| 1st | $(100), (010), (\bar{1}10),$ $(\bar{1}00)$ $(0\bar{1}0), (1\bar{1}0)$ | $(01\bar{1}0), (\bar{1}100), (\bar{1}010)$ $(0\bar{1}10), (1\bar{1}00), (10\bar{1}0)$ | $\{10\bar{1}0\}$ | 7.347 | |
| 2nd | $(110), (\bar{2}10), (1\bar{2}0),$ $(\bar{1}20)$ $(\bar{1}\bar{1}0), (2\bar{1}0)$ | $(11\bar{2}0), (\bar{2}110), (1\bar{2}10)$ $, (\bar{1}210)$ $(\bar{1}\bar{1}20), (2\bar{1}\bar{1}0)$ | $\{11\bar{2}0\}$ | 7.289 | |
| 3rd | $(201), (021), (\bar{2}21),$ $(\bar{2}01),$ $(0\bar{2}1), (2\bar{2}1), (20\bar{1}),$ $(02\bar{1}),$ $(\bar{2}2\bar{1}), (\bar{2}0\bar{1}), (0\bar{2}\bar{1}),$ $(2\bar{2}\bar{1})$ | $(20\bar{2}1), (02\bar{2}1), (\bar{2}\bar{2}01)$ $, (\bar{2}021)$ $(0\bar{2}21), (2\bar{2}01),$ $(20\bar{2}\bar{1}), (02\bar{2}\bar{1})$ $(\bar{2}20\bar{1}), (\bar{2}02\bar{1}), (0\bar{2}2\bar{1})$ $, (2\bar{2}0\bar{1})$ | $\{20\bar{2}1\}$ | 7.318 [ Average of the values of $a$ of 1st and 2nd rings] | 5.281 [Obtained by using $a$ =7.318 μm] |
| 4th | $(211) (\bar{1}31), (\bar{3}21),$ $(\bar{2}1\bar{1}),$ $(1\bar{3}1), (3\bar{2}1), (21\bar{1}),$ $(\bar{1}3\bar{1}),$ $(\bar{3}2\bar{1}), (\bar{2}\bar{1}\bar{1}), (1\bar{3}\bar{1}),$ $(3\bar{2}\bar{1})$ | $(21\bar{3}1), (\bar{1}\bar{3}21), (\bar{3}211),$ $(\bar{2}1\bar{3}1)$ $(1\bar{3}21), (3\bar{2}\bar{1}1), (21\bar{3}\bar{1}),$ $(\bar{1}\bar{3}2\bar{1})$ $(\bar{3}21\bar{1}), (\bar{2}1\bar{3}\bar{1}), (1\bar{3}2\bar{1}),$ $(3\bar{2}\bar{1}\bar{1})$ | $\{21\bar{3}1\}$ | 7.421 [Obtained by using $c$ =5.281 μm] | |